\documentclass[12pt]{article}

\usepackage{amsmath}
\usepackage{amssymb}
\usepackage{amsfonts}
\usepackage{lscape}

\begin{document}

\fontsize{12}{6mm}\selectfont
\setlength{\baselineskip}{2em}

$~$\\[.35in]
\newcommand{\dss}{\displaystyle}
\newcommand{\raro}{\rightarrow}
\newcommand{\be}{\begin{equation}}

\def\sech{\mbox{\rm sech}}
\def\sn{\mbox{\rm sn}}
\def\dn{\mbox{\rm dn}}
\thispagestyle{empty}

\begin{center}
{\Large\bf Hamiltonians for the Quantum Hall Effect}  \\    [2mm]
{\Large\bf on Spaces with Non-Constant}  \\    [2mm]
{\Large\bf Metrics}   \\   [2mm]
\end{center}

\vspace{1cm}
\begin{center}
{\bf Paul Bracken}                        \\
{\bf Department of Mathematics,} \\
{\bf University of Texas,} \\
{\bf Edinburg, TX  }  \\
{78541-2999}
\end{center}

\vspace{3cm}
\begin{abstract}
The problem of studying the quantum Hall effect on
manifolds with non constant metric is addressed.
The Hamiltonian on a space with hyperbolic metric
is determined, and the spectrum and eigenfunctions are
calculated in closed form. The hyperbolic disk is also 
considered and some other applications of this 
approach are discussed as well.
\end{abstract}

\vspace{2mm}
\vspace{2mm}

\newpage
\begin{center}
{\bf 1. Introduction.}
\end{center}

The problem of investigating the quantum Hall effect (QHE)
on different types of manifold has been of great interest
recently. Some of the previous work in this area originated
in the course of generalizing the Hall current from the
$SO (3)$ two-sphere $S^2$ to the four-sphere $S^4$
of the invariance group $SO(5)$ {\bf [1]}. Quantum Hall droplets
have been considered as well on complex projective spaces
$CP^d$ {\bf [2]}. The idea all along has been to
generalize the Landau problem on different types of higher
dimensional spaces. This kind of work combines the study
of several important areas in mathematics and physics.

Zhang and Hu {\bf [3]} considered the Landau problem
for charged fermions on $S^4$ under the influence of a
background magnetic field which is related to the standard 
$SU (2)$ instanton. In ordinary quantum Hall effects, a
droplet of fermions occupying a certain region
behaves as an incompressible fluid, a characteristic 
property of the QHE, the low energy excitations being
area-preserving deformations which behave as massless
chiral bosons. Some work on chiral boson theories
related to the QHE has been described in {\bf [4]}.
Now $S^4$ may be considered since the edge excitations
could lead to higher spin massless fields, in particular,
the graviton. The original work of Zhang and Hu {\bf [5]} has been 
continued to include the QHE on even-dimensional
complex projective spaces $CP^k$, for example.
This is interesting since one can obtain incompressible
droplet states by coupling the fermions to a $U (1)$
background field.

The Landau problem has been of fundamental importance to
the QHE from the beginning {\bf [6,7]}. The main aim here is to
consider generalizations of the QHE to spaces with
nonflat metrics by considering Hamiltonians 
of the Laplace-Beltrami form. For example, the spectrum
of the Laplacian on the ball $B^d$ in different
dimensions has been investigated. Of interest here will be
the generalization of the Landau problem on the plane 
to the case of spaces with non-constant metrics.
The Hall conductance can be thought of as a type
of curvature, and it would be useful to understand
relationship more thoroughly and in other frameworks.
We begin by introducing a classical model of a charged 
particle in a magnetic field on the hyperbolic plane {\bf [8]}.
The metric in this case is diagonal but non-constant.
Next the quantum version of the Landau problem on the 
plane will be reviewed for the sake of completeness,
and most of the important results will be derived {\bf [9]}.
The quantized version for the related problem under
a diagonal but non-constant hyperbolic metric will be formulated next.
The Hamiltonian can be derived by making use of the 
Laplace-Beltrami operator for a particle of mass $m$
on a manifold with metric $g_{ij}$ under a monopole
field. A separation of variables solution can be obtained 
and the associated eigenvalues and eigenfunctions for this
differential equation can be written down explicitly.
The same problem can then be formulated from the Lie
algebraic point of view as well. The Hamiltonian can be
expressed in terms of the relevant Casimir operator,
and the energies can be obtained in this way as well.
Some interesting observations with regard to the
operator ordering question for the Hamiltonian as well 
as the underlying symmetry vector fields can be made.
Finally, this will be repeated for the case of 
formulating a Hamiltonian for the hyperbolic disk.

\begin{center}
{\bf 2. Classical Lagrangian Formulation in a Hyperbolic Space.}
\end{center}

Before investigating the quantum dynamics of a charged 
particle under the influence of a magnetic field
in a space with a non-constant metric, it is worth
reviewing the classical formulation of the dynamics
of a charged particle in a space with a hyperbolic metric.
Consider dynamics then on the upper half of the
complex plane, the Poincar\'e plane defined by,
$$
{\mathbb H} = \{ z = x + i y  \in \mathbb C , y >0 \}.
\eqno(2.1)
$$
The metric on this space can be written as
$$
ds^2 = \frac{a^2}{y^2} ( dx \otimes dx +
dy \otimes dy ).
\eqno(2.2)
$$
The scalar curvature for this space is a negative
constant $K =- 2/a^2$. A constant magnetic field
can be obtained from a vector potential in the plane
given by
$$
{\bf A} = ( - \frac{\beta}{y}, 0),
\qquad
\beta = B a^2.
\eqno(2.3)
$$
The quantity $\beta$ defined in (2.3)
is frequently referred to as the rescaled field.

The Euler-Lagrange equations which determine the
classical trajectories of a charged particle in
the field generated by ${\bf A}$ can be determined
from the Lagrangian
$$
L = \frac{a^2}{y^2} ( \dot{x}^2 + \dot{y}^2 ) - \beta \frac{\dot{x}}{y}.
\eqno(2.4)
$$
The differentiation in (2.4) is with respect to the 
time variable. The Hamiltonian is determined from $L$  
by means of
$$
H = \sum_{r=1}^{N} p_r \dot{q}_{r} - L,
\eqno(2.5)
$$
where
$$
p_x = p_1 = \frac{\partial L}{\partial \dot{x}}
= \frac{2 a^2}{y^2} \dot{x} - \frac{B}{y} a^2,
\qquad
p_y = p_2 = \frac{\partial L}{\partial \dot{y}}
= \frac{2 a^2}{y^2} \dot{y}.
\eqno(2.6)
$$
Solving this pair individually for $\dot{x}$ and $\dot{y}$,
the results are substituted into (2.5) to obtain
$$
H = p_x \dot{x} + p_y \dot{y}
- \frac{a^2}{y^2} ( \dot{x}^2 + \dot{y}^2 ) + \beta \frac{\dot{x}}{y}
$$
$$
= p_x ( \frac{y^2}{2 a^2} p_x + \frac{y}{2} B)
+ p_y \frac{y^2}{2 a^2} p_y - \frac{a^2}{y^2}
( \frac{y^2}{2 a^2} p_x + \frac{y}{2} B)^2
- \frac{y^2}{4 a^2} p_y^2 + \frac{\beta}{y}
( \frac{y^2}{2 a^2} p_x + \frac{y}{2} B)
$$
$$
= \frac{1}{4 a^2} \{ y^2 ( p_x^2 + p_y^2 )+ 2 \beta y p_x + \beta^2 \}.
\eqno(2.5)
$$
Constants of the motion are determined by Noether's
theorem and are given by
$$
L_1 = x p_x + y p_y,
$$
$$
L_2 = p_y,
\eqno(2.8)
$$
$$
L_3 = (y^2 -x^2 ) p_x
- 2 xy p_y + 2 \beta y.
$$
Classically the relative order of each of the factors
in the $L_j$ in (2.8) is irrelevant. Thus, by direct
calculation, it follows that
$$
L_2 L_3 + L_1^2 = y^2 ( p_x^2 + p_y^2 ) + 2 \beta y p_x.
$$
By comparison with (2.7), it follows that
$$
H = \frac{1}{4 a^2} ( L_2 L_3 + L_1^2 + \beta^2).
\eqno(2.9)
$$
By resorting to Hamiltonian (2.7), Hamilton's equations
can be formulated directly as
$$
\begin{array}{cc}
\dot{x} = \dss \frac{\partial H}{\partial p_x} =
\dss \frac{1}{2 a^2} ( y^2 p_x + \beta y),  &
\dot{y} = \dss \frac{\partial H}{\partial p_y} =
\dss \frac{y^2}{2 a^2} p_y,   \\
   &   \\
\dot{p}_{x} = - \dss \frac{\partial H}{\partial x} =0, &
\dot{p}_y =- \dss \frac{\partial H}{\partial y}
= \dss \frac{1}{2 a^2}
( y ( p_x^2 + p_y^2 ) + \beta p_x).
\end{array}
\eqno(2.10)
$$
The Poisson bracket of any two classical functions
$F$ and $G$ is given by
$$
\{ F, G \} = \sum_{r=1}^{N} ( \frac{\partial F}{\partial q^r}
\frac{\partial G}{\partial p_r}
- \frac{\partial F}{\partial p_r} \frac{\partial G}{\partial q^r} ).
\eqno(2.11)
$$
By straightforward differentiation of the $L_j$ in
(2.8), it can be verified that, with the canonical Poisson 
bracket, the $L_j$ given in
(2.8) generate an $sl (2, \mathbb R)$ Lie algebra 
$$
\{ L_1 , L_2 \} = L_2,
\qquad
\{ L_1 , L_3 \} = - L_2 ,
\qquad
\{ L_2, L_3 \}  = 2 L_1.
\eqno(2.12)
$$
Eliminating $p_x$ and $p_y$ from (2.8), the classical
path can be determined, and with respect
to the Euclidean plane, it describes a circle.

\begin{center}
{\bf 3. Quantum Problem in the Plane.}
\end{center}

The dynamics of a charged particle in the plane
under the influence of an external uniform magnetic
field $B$ which is oriented at right angles to the plane
will be studied first. This will provide a setting in which 
to introduce the QHE as well. It will be useful
here to frequently use complex coordinates $z$ and $\bar{z}$
defined by $z = x + i y$ and $\partial = \partial/ \partial z$
and $\bar{\partial} = \partial / \partial \bar{z}$.
The gauged Hamiltonian is usually written in the form
$$
H = \frac{1}{2m} ( {\bf p} - \frac{e}{c} {\bf A} )^2.
\eqno(3.1)
$$
In the symmetric gauge and formulated in terms of complex $z$ and $\bar{z}$,
(3.1) can be written
$$
H = - \frac{2 \hbar^2}{m} \bar{\partial} \partial
+ \frac{m \omega_0^2}{8} | z|^2 - \frac{\hbar \omega_c}{2}
( z \partial - \bar{z} \bar{\partial}),
\eqno(3.2)
$$
where $\omega_c = e |B|/mc$ is the cyclotron frequency.

Introduce two quantum operators $a$ and $a^{\dagger}$
defined by,
$$
a = -2i \sqrt{\frac{\hbar}{2m \omega_c}} (\bar{\partial} 
+ \frac{m \omega_c}{4 \hbar} z),
\qquad
a^{\dagger} = -  2i \sqrt{\frac{\hbar}{2 m \omega_c}}
(\partial - \frac{m \omega_c}{4 \hbar} \bar{z} ).
\eqno(3.3)
$$
Using these, we calculate
$$
a a^{\dagger} - a^{\dagger} a = 1,
\eqno(3.4)
$$
and 
$$
\frac{\hbar \omega_c}{2} ( a a^{\dagger} +
a^{\dagger} a) = - \frac{2 \hbar^2}{m} \partial
\bar{\partial}
- \frac{\hbar \omega_c}{2} ( z \partial - \bar{z} \bar{\partial})
+ \frac{1}{8} m \omega_c^2 |z|^2.
\eqno(3.5)
$$
Therefore, in terms of the operators defined by (3.3),
we can write
$$
H = \frac{1}{2} \hbar \omega_c ( 2 a^{\dagger} a + 1).
\eqno(3.6)
$$
Now $a^{\dagger} a$ can be interpreted as a number
operator, thus defining a number basis $| n >$, the energy
spectrum of $H$ is given by
$$
E_n = ( n + \frac{1}{2} ) \hbar \omega_c,
\qquad
n=0,1,2, \cdots.
\eqno(3.7)
$$
The Landau levels $E_n$ are degenerate with respect
to the center of Larmor's circular orbits.
Corresponding eigenfunctions are obtained from the
ground state eigenfunction as
$$
| n > = \frac{(a^{\dagger})^n}{\sqrt{n!}} |0>,
\eqno(3.8)
$$
where the ground state $|0>$ obeys the equation
$$
a | 0 > = 0.
\eqno(3.9)
$$
Substituting $a$ from (3.3), this implies a first
order equation for the ground state function
given by
$$
( \frac{\partial}{\partial \bar{z}} + \frac{m \omega_c}{4 \hbar} z )
\psi_0 ( z, \bar{z} ) = 0.
\eqno(3.10)
$$
This equation has the general solution
$$
\psi_0 ( z, \bar{z} ) = f(z) \exp ( - \frac{|z|^2}{4 z_0^2}),
\eqno(3.12)
$$
where $z_0 = \sqrt{ \hbar c/ e B}$.

The Hamiltonian in (3.2) can be generalized to
a many-particle system described by the total
Hamiltonian
$$
 H = \sum_{i=1}^{N} \{
- \frac{2 \hbar^2}{m} \bar{\partial}_{i} \partial_{i}
- \frac{\hbar \omega_c}{2} ( z_i \partial_i
- \bar{z}_{i} \bar{\partial}_{i} )
+ \frac{ m \omega_{c}^2}{8} | z_i|^2 \}.
\eqno(3.13)
$$
Suppose we consider $N$-particles in the lowest
Landau level, which means that the quantum numbers
$n_i = 0$ with $i=1,2 \cdots, N$ and each $n_i$ 
corresponds to the spectrum (3.8). The total
wavefunction can be written in terms of the
Slater determinant
$$
\Psi ( z, \bar{z} ) = \epsilon^{i_1 \cdots i_N}
z_{i_1}^{n_1} \cdots z_{i_N}^{n_{N}} \exp
( - \sum_{i=1}^{N} \frac{|z_i|^2}{4 z_0^2} ),
\eqno(3.14)
$$
where $\epsilon^{i_1 \cdots i_N}$ is the fully
antisymmetric tensor and the $n_i$ are integers.
As a Vandermonde determinant, we have
$$
\Psi_{1} ( z, \bar{z} ) = K \prod_{i,j}
( z_i - z_j) \exp ( - \sum_{i=1}^N \frac{|z_i|^2}{4 z_0^2}).
\eqno(3.15)
$$
This can be compared with the Laughlin wavefunction given by
$$
\Psi_m ( z , \bar{z} ) = K_{m} \prod_{i,j}
( z_i - z_j )^{m} \,
\exp ( - \sum_{i=1}^{N} \frac{|z_i|^2}{4 z_0^2} ).
\eqno(3.16)
$$
This matches (3.15) when $m=1$ and is a good ansatz
to describe the fractional QHE at the filling factor
$\nu = 1/m$.

Let us mention that the filling factor in units
such that $h e/c$ is one can be written as
$$
\nu = \frac{2 \pi {\cal N}}{B},
\eqno(3.17)
$$
where ${\cal N}$ is the density of particles
$$
{\cal N} = \frac{N}{S},
$$
and $S$ is the plane surface area. In the QHE,
$\nu$ in (3.17) must be quantized and reads
$$
\nu = \frac{N}{N_{\phi}}.
\eqno(3.18)
$$
This can be either an integer or fractional depending
on which kind of effect is involved and $N_{\phi}$
is the quantum flux number. Thus, the
magnetic flux is quantized here.

\begin{center}
{\bf 4. The Quantum Problem in a Hyperbolic Geometry.}
\end{center}

A single particle quantum Hamiltonian can be
developed for the Poincar\'e half-plane. It can be 
derived from the electromagnetically gauged
form of the Laplace-Beltrami operator, such that
the contribution of the vector potential has
been included. In this section, units are taken
such that $\hbar = c = e=1$.
If the particle has a mass $m$ on a
space with metric $g_{ij}$, the operator can
be written as
$$
H^{LB} = \frac{1}{2m \sqrt{g}} ( {\bf p} - {\bf A})_{i}
( \sqrt{g} g^{ij} ) ( {\bf p } - {\bf A})_{j},
\eqno(4.1)
$$
where $g$ in (4.1) is the determinant of the metric,
$g^{ij}$ the inverse of $g_{ij}$,
and we follow de Witt's prescription {\bf [10]} such that the
covariant derivative ${\bf p}$ contains a 
contribution which is directly related to the
metric. The metric for the space which is of
interest here is given by (2.1), and the gauge
is fixed by taking the vector potential in the
plane to have the form (2.3). From the form
of the metric, it is clear that $\sqrt{g} =
a^2/ y^2$ and the inverse metric has elements which
are the reciprocals of those in $g_{ij}$.
When the metric has diagonal form, the
Laplace-Beltrami operator takes the classical
form
$$
H^{LB} = \frac{y^2}{2m a^2} ( P_1^2 + P_2^2 ),
\eqno(4.3)
$$
where the momenta $P_j$ are gauged with the
electromagnetic contributions due
to a nontrivial vector potential, which
can be written 
$$
P_j = p_j - A_j.
\eqno(4.4)
$$
When $H^{LB}$ is quantized, the form for $H^{LB}$
given in (4.3) could be adopted, or the operators
$y$ could be placed in a different order
$$
H^{LB} = \frac{1}{2m a^2} \, y ( P_1^2 + P_2^2) y,
\eqno(4.5)
$$
or even with the factor of $y^2$ placed entirely
to the far right of the operator. These different 
cases just correspond  to the usual operator ordering 
ambiguities  which arise during quantization.
Some remarks related to this will be made later.

To finish the canonical quantization procedure 
following de Witt's prescription, the momentum
operators $p_j$ have to be constructed, and these
contain a contribution which is related to the metric
determinant $g$, and are given as follows
$$
\begin{array}{c}
p_1 = p_x = -i (\partial_x + \frac{1}{2} \partial_x \ln \sqrt{g} )
= -i \partial_x,    \\
   \\
p_2 = p_y = -i ( \partial_y + \frac{1}{2} \partial_y \ln \sqrt{g} )
= -i \partial_y + \dss \frac{i}{y}.
\end{array}
\eqno(4.6)
$$
Substituting $p_j$ given in (4.6) into the
Hamiltonian (4.5), we obtain that
$$
H^{LB} = \frac{1}{2m a^2} \, y \{ (-i \partial_x +
\frac{\beta}{y} )^2 + (-i \partial_{y} + \frac{i}{y})^2 \} y,
\eqno(4.7)
$$
and $\beta$ is defined in (2.3).
Expanding the terms in this operator out in full,
we find
$$
\begin{array}{c}
y ( -i \partial_x + \dss \frac{\beta}{y} )^2 y
= - y^2 \partial_x^2 -2i \beta y \partial_x + \beta^2,  \\
   \\
y ( -i \partial_y + \dss \frac{i}{y} )(-i \partial_{y} y +i)
=- iy (-i \partial_y + \dss \frac{i}{y})y \partial_y 
= - y^2 \partial_y^2.
\end{array}
\eqno(4.8)
$$
Therefore, the Hamiltonian (4.5) takes the form,
$$
H^{LB} = \frac{1}{2m a^2} \{ -y^2 (\partial_x^2 
+ \partial_y^2) - 2 i \beta y \partial_{x} + \beta^2 \},
\eqno(4.9)
$$
and the associated eigenvalue problem takes the form
$$
H^{LB} \Psi = E \Psi.
\eqno(4.10)
$$
Just as in the case of (4.5), 
the three constants of the motion $L_j$ in (1.6)
are also determined up to ordering ambiguities as
well. The exact form of the operators can be
established in this case by using them to establish
a particular Lie algebra structure such that the
associated Casimir operator matches the Hamiltonian 
(4.9) up to constant terms. 
In this case remarkably, combining these approaches
seems to be sufficient to eliminate the ordering
ambiguities, that is, the ordering given in the quantum
form of the $L_j$ is the one which matches (4.5), and the
other cases for the $L_j$ and $H^{LB}$ with $y$ in different
orientations will not coincide under this procedure.
Consider the following
form for the operators $L_j$,
$$
L_1 = -i \partial_x \, x + y ( -i \partial_y + \frac{i}{y}),
$$
$$
L_2 =-i \partial_x,
\eqno(4.11)
$$
$$
L_3 =-i  \, \partial_x \, (y^2 - x^2)
-2 xy (-i  \partial_y + \dss \frac{i}{y} ) + 2 \beta y.
$$
Moving all the operator terms to the right-hand side
of the variables by expanding out, these become
$$
L_1 =-i ( x \, \partial_x + y \, \partial_y ), 
$$
$$
L_2 = -i \partial_x,
\eqno(4.12)
$$
$$
L_3 =-i (y^2 -x^2 ) \, \partial_x + 2 i xy \partial_y
+ 2 \beta y.
$$
The calculations which can easily be carried
out by using MAPLE {\bf [11]} and has been done this way. 
It is found that the
operators in (4.12) satisfy the following 
commutation relations
$$
[ L_1 , L_2 ] = i L_2,
\qquad
[L_1 , L_3 ] = -i L_3,
\qquad
[L_2 , L_3 ] = 2 i L_1.
\eqno(4.13)
$$
The following combinations of the operators $L_j$ 
can be written down
$$
J_0 = \frac{1}{2} (L_2 - L_3),
\qquad
J_1 = \frac{1}{2} (L_2 +L_3),
\qquad
J_2 = L_1.
\eqno(4.14)
$$
If the $L_j$ satisfy (4.13), it is easy to show 
using the properties of the commutator that the
$J_i$ defined in (4.14) satisfy the canonical form of the $su (1,1)$
algebra described by the commutators {\bf [12,13]}
$$
[ J_0 , J_1 ] = i J_2,
\qquad
[ J_0, J_2 ] =- i J_1,
\qquad
[ J_1, J_2 ] = - i J_0.
\eqno(4.15)
$$
The Casimir operator which corresponds to these operators
under this algebra is given by
$$
C = J_0^2 - J_1^2 - J_2^2.
\eqno(4.16)
$$
Using (4.14), this can be written in terms of the 
$L_j$ as follows
$$
C = - L_2 L_3 - L_1^2 + i L_1.
\eqno(4.17)
$$
Again, by direct calculation, it can be verified
using (4.12) that
$$
-C = L_2 L_3 + L_1^2 -i L_1 
= - y^2 (\partial_x^2 + \partial_y^2 ) - 2 i \beta y \partial_x.
$$
Hence in terms of the Casimir operator $C$, the
Hamiltonian (4.9) can be written in the equivalent form
$$
H^{LB} = \frac{1}{2m a^2} (-C + \beta^2).
\eqno(4.18)
$$
The spectrum of $H^{LB}$ can be obtained from the
representation theory that corresponds to the operator $C$.
Consider a unitary, irreducible representation of the
group as eigenstates of $C$ as well as the compact
generator $J_0$. Let us choose a basis $| j, m >$ such that
$$
C | j,m > = j ( j+1) | j,m >,
\eqno(4.18)
$$
such that $m$ is an eigenvalue of $J_0$,
$$
J_0 | j,m > = m | j,m>.
\eqno(4.20)
$$
Using (4.18), the effect of the Hamiltonian on
$|j,m>$ can be determined  
$$
H^{LB} |j,m > = \frac{1}{2 m a^2} (\beta^2 - j (j+1) ) | j,m >.
\eqno(4.21)
$$
Substituting into eigenvalue problem (4.10), 
and taking $j=l-b$ where $l$ is an integer, the
energy eigenvalues can be expressed in the form,
$$
E_{\beta,l} = \frac{1}{2m a^2} ( \beta^2 + \frac{1}{4}
- ( l - \beta + \frac{1}{2})^2 ).
\eqno(4.22)
$$
It will be useful to compare the result in (4.22)
with that obtained by solving the eigenvalue problem (4.10)
in differential form.
Consider now (4.10)  with $H^{LB}$ given by (4.9).
Let us consider obtaining a class of solutions
which have a separation of variables form
$$
\Psi (x,y) = \alpha (x) \varphi (y).
\eqno(4.23)
$$
Substituting (4.23) into (4.10), the differential equation 
takes the form
$$
- y^2 ( \frac{\alpha_{xx}}{\alpha} + \frac{\varphi_{yy}}{\varphi})
- 2 i \beta y \frac{\alpha_{x}}{\alpha} + \beta^2 - 2m a^2 E =0.
\eqno(4.24)
$$
To decouple the $x$ and $y$ variables in (4.24),
let us require that $\alpha_x =-i c \alpha$, where $c$
is a real, positive constant hence we take
$$
\alpha ( x ) = e^{-i c x}.
\eqno(4.25)
$$
The appearance of the imaginary unit in the exponential
reduces the equation entirely to real form. Substituting
(4.25) into (4.24), we obtain an equation in terms of $\varphi$
$$
\frac{\varphi_{yy}}{\varphi} - c^2 + \frac{2 \beta c}{y}
+ \frac{1}{y^2} ( - \beta^2 + 2m E) = 0.
\eqno(4.26)
$$
Introducing the variable to $s= 2 c y$ and setting
$- \beta^2 + 2 m E = \frac{1}{4} - n^2$, equation
(4.26) takes the form of a Whittaker equation
$$
\frac{\varphi''}{\varphi} - \frac{1}{4} + \frac{\beta}{s} 
+ \frac{1}{s^2} ( \frac{1}{4} - n^2) = 0,
\eqno(4.27)
$$
such that the energies are related to $n$ through
$$
E_{n} = \frac{1}{2 m a^2} ( \frac{1}{4} - n^2 + \beta^2).
\eqno(4.28)
$$
The general form for the solutions to (4.27) can be
written in terms of the confluent hypergeometric
function 
$$
M_{\beta,n} (s) = e^{-s/2} s^{1/2 +n} \, _{1}F_{1}
( \frac{1}{2}+n- \beta ; 1+2n;s),
\qquad
M_{\beta, -n} (s) =  e^{-s/2} s^{1/2 -n} \,  _{1}F_{1}
( \frac{1}{2}-n- \beta; 1-2n ;s).
\eqno(4.29)
$$
If $n$ is taken such that $n = \beta - l - \frac{1}{2}$,
where $l$ is chosen to be an integer such that
$0 \leq l < \beta - \frac{1}{2}$, the confluent
hypergeometric function in $M_{\beta,n}$ will
truncate to the form of a Laguerre polynomial {\bf [14,15]}.
Moreover, $M_{\beta,n} (2cy)$ is defined at $y=0$ and
square integrable in $y$ on the domain (2.1).
The connection between the hypergeometric function and
Laguerre polynomial is provided by
$$
L_{n}^{(\tau)} (z) = \frac{(\tau + 1)_{n}}{n!}
\, _{1}F_{1} (-n ; \tau+1;z).
$$
Substituting this into $M_{\beta,n} (s)$ in (4.29),
the eigenfunctions of (4.10), up to a normalization
constant can be written in the form
$$
\Psi (x,y) = {\cal N} e^{-icx - cy} y^{\beta -l}
L_{l}^{2 \beta -2l-1} ( 2cy),
\eqno(4.30)
$$
and the energies are given by (4.28) with
$n = \beta -l -1/2$ as,
$$
E_{\beta,l} = \frac{1}{2m a^2} ( \beta^2
+ \frac{1}{4} - ( l - \beta + \frac{1}{2})^2 ).
\eqno(4.31)
$$
These results for $E_{\beta,l}$ can be compared
with the expression given in (4.22).

Note that (4.9) can be written in complex form as
well
$$
H^{LB} = \frac{1}{2m a^2}
 \{ (z - \bar{z} )^2 \bar{\partial} \partial - 
\beta ( z - \bar{z} ) ( \partial + \bar{\partial} ) 
+ \beta^2 \}.
\eqno(4.32)
$$
For equal mass particles, (4.32) can be generalized 
to the case of a many-particle system as was done
in (3.13) to give the Hamiltonian
$$
H^{LB} = \frac{1}{2m a^2} \sum_{i=1}^{N}
[ (z_i - \bar{z}_i ) ^2 \bar{\partial}_i \partial_i - \beta ( z_i - \bar{z}_i)
( \partial_i + \bar{\partial}_i ) + \beta^2].
\eqno(4.33)
$$

\begin{center}
{\bf 5. Results for Other Geometries and Conclusions.}
\end{center}

A Hamiltonian will be developed using the method
described in the last section for a system on the
hyperbolic disk ${\mathbb B}^1_{\rho}$, which is defined to be
$$
{\mathbb B}^1_{\rho} = \{ w = x + iy \in {\mathbb C}
| |w|^2 < \rho^2 \},
\eqno(5.1)
$$
which carries the Bergman-K\"{a}hler metric
$$
ds^2 = ( 1 - \frac{|w|^2}{\rho^2} )^{-2}
( dx \otimes dx + dy \otimes dy ),
\eqno(5.2)
$$
where $|w|^2 = x^2 + y^2$, so the metric is again diagonal.
Define the function
$$
\phi (x,y) = 1 - \frac{x^2 + y^2}{\rho^2},
\eqno(5.3)
$$
and clearly $\sqrt{g} = \phi^{-2}$ in this case.
To write a Hamiltonian on ${\mathbb B}^1_{\rho}$,
we work out (4.1) with the metric (5.2) and take a
vector potential with two components which is defined
everywhere on (5.1). For $x$ and $y$ such that $|w|^2 < \rho$,
then with $B$ the magnetic field in a symmetric gauge let 
$$
{\bf A} = B ( y, -x ).
\eqno(5.4)
$$
Using the definitions in (4.6), the vector potential
is specified by
$$
p_x = p_1 =- i (\partial_x + \frac{2x}{\rho^2 \phi} ),
\qquad
p_y = p_2 = -i (\partial_y + \frac{2y}{\rho^2 \phi} ).
\eqno(5.5)
$$
Therefore, substituting (5.4) into (4.1), we have
$$
H = \frac{1}{2m} \phi^2 (( -i \partial_x - \frac{2ix}{\rho^2 \phi}
- By)^2  + ( -i \partial_y - \frac{2iy}{\rho^2 \phi} + Bx)^2)
$$
$$
= \frac{\phi}{2m} \{ - \phi ( \partial_x^2 + \partial_y^2)
- \frac{4}{\rho^2} ( x \partial_x + y \partial_y)
+ 2 i B \phi ( y \partial_x - x \partial_y) + B^2 \phi
- \frac{4}{\rho^2} ( 1 + \frac{2 |w|^2}{\rho^2 \phi} ) \}.
\eqno(5.6)
$$
As a last application of geometrical methods to generating
Hamiltonians for this area, consider the approach proposed
by Haldane {\bf [16]} to overcoming the symmetry problem in the
Laughlin theory of the fractional QHE described by
wavefunction (3.16) at the filling fraction $\nu= 1/m$.
To phrase the problem more precisely, (3.16) is rotationally
invariant due to angular momenta, but it is not translationally
invariant. By considering particles living on a two-sphere
in a magnetic monopole, Haldane formulated a theory
that has all the symmetries and generalizes the
Laughlin proposal.

The link with the two-sphere $S^2$ can be done by means of
the following approach. First $S^2$ can be realized on the
disk
$$
\partial {\mathbb B}^1_{\rho} = S^2 = \{ w \in \mathbb C |
|w| = \rho \},
$$
which is the boundary of ${\mathbb B}^1_{\rho}$, and the
basic features of ${\mathbb B}_{\rho}^1$ lead to those of
$S^2$. The space $H$ is invariant on the symmetric
space $SU (1,1) / U(1)$ and the projective space
$CP^1$ can be obtained as
$$
CP^1 = SU (2) / U (1).
$$
The $S^2$ can be regarded as an analytic continuation
of $SU (1,1)$ to $SU (2)$. This suggests that the
spectrum obtained on ${\mathbb B}^1_{\rho}$ is
similar to the Landau problem on the sphere, except
that the eigenfunctions should be invariant under
the group $U (1)$.

The $CP^1$ expression shows that functions on $S^2$
can be thought of as functions of $SU (2)$,
invariant under the $U (1)$ subgroup. A basis of functions 
for $SU (2)$ is given by the Wigner $D$-functions,
and a basis for functions on $S^2$ is given by the $SU (2)$
Wigner functions $D^{(j)}_{L_3 R_3} (g)$ with trivial
right $U (1)$ action, that is, the $U (1)_R$ charge
$R_3 =0$. Derivatives on $S^2$ can be identified as $SU (2)$
right rotations $SU (2)_R$, which satisfy an $SU(2)$ algebra.
Consequently, covariant derivatives $D_{\pm}$ can be 
written as
$$
D_{+} = \frac{1}{\rho} L_{2},
\qquad
D_{-}  = \frac{1}{\rho} L_{3}.
\eqno(5.7)
$$
The operators in (5.7)  must satisfy the commutator bracket
$$
[ D_{+} , D_{-} ] = - \frac{B}{2}.
\eqno(5.8)
$$
Defining $k = B \rho^2 /2$ and substituting (5.7) into (5.8),
the commutator and covariant derivatives 
fix the eigenvalue of $L_1$ to be
$$
L_{1} = \frac{i}{2} k.
\eqno(5.9)
$$
A Hamiltonian can be written down in terms of the $D_{\pm}$
as follows,
$$
H = - (D_{+} D_{-} + D_{-} D_{+} )
= - \frac{1}{\rho^2} ( L_2 L_3 + L_3 L_2)
= - \frac{2}{\rho^2} ( L_2 L_3 -i L_1).
\eqno(5.10)
$$
This can be written in terms of the Casimir
operator (4.16) using $L_2 L_3 =- C +i L_1 - L_1^2$
in the following way
$$
H = \frac{2}{\rho^2} ( C + L_1^2).
\eqno(5.11)
$$
Using the representation theory in which the
eigenvalues of $C$ are $j (j+1)$ and $j$ is taken
to be $ j = l - k/2$, the associated spectrum of
(5.11) is
$$
E_{l} = \frac{2}{\rho^2} [ ( l - \frac{k}{2})
( l - \frac{k}{2} +1) - \frac{k^2}{4} ].
\eqno(5.12)
$$

It is clear that this geometric approach yields
useful results both in the nature of fundamental
evolution equations and predictions for the
trajectories in the classical case, as well as
predictions for the energy spectrum and wave 
functions in the quantum problem. It will also be of interest
to extend the work in this study to problems
in higher dimensions, and to consider the 
existence of other classes of solutions to
the equations, such as soliton solutions.

\newpage
\begin{center}
{\bf References.}
\end{center}

\noindent
$[1]$ S. C. Zhang and J. P. Hu, Science {\bf 294}, 823 (2001).  \\
$[2]$ D. Karabali and V. P. Nair, Nucl. Phys. {\bf 679}, 427 (2004),
Nucl. Phys. {\bf B 641} 533 (2002).  \\
$[3]$ S. C. Zhang and J. P. Hu, cond-mat/0110572.   \\
$[4]$ P. Bracken, Can. J. Phys. {\bf 79} 1121 (2001).     \\
$[5]$ J. P. Hu and S. C. Zhang, Phys. Rev. {\bf B 66}, 125301 (2002).  \\
$[6]$ Z. F. Ezawa, Quantum Hall Effects, (World Scientific,
Singapore, 2000).   \\
$[7]$ A. Khare, Fractional Statistics and Quantum Theory, (World Scientific,
Singapore, 2005).  \\
$[8]$ A. Comtet, Ann. of Phys. {\bf 173}, 185 (1987).  \\
$[9]$ A. Jellal, Nucl. Phys. {\bf B725}, 554 (2005), R. Iengo and D. Li,
Nucl. Phys. {\bf B 413}, 735 (1994), F. Chandelier, Y. Georgelin,
T. Masson, J.-C. Wallet, Ann. of Phys. {\bf 314} 476 (2004).  \\
$[10]$ B. S. De Witt, Rev. Mod. Phys. {\bf 29 (3)}, 377 (1957).  \\
$[11]$ B. W. Char, K. O.Geddes, G. H. Gonnet, B. Leong,
M. Monagan, S. Watt, Maple Library Reference Manual, (Waterloo Maple 
Publishing, 1991).  \\
$[12]$ J. Chen, J. Ping and F. Wang, Group Representation Theory
for Physicists, (World Scientific, Singapore, 2002).  \\
$[13]$ A. O. Barut and R. Raczka, Theory of Group Representations
and Applications, (World Scientific, Singapore, 1986).   \\
$[14]$ G. Andrews, R. Askey and R. Roy, Special Functions,
(Cambridge University Press, Cambridge, 19990.  \\
$[15]$ Y. L. Luke, The Special Functions and their Approximation, Vol 1,
(Academic Press, New York, 1969).   \\
$[16]$ F. D. Haldane, Phys. Rev. Lett. {\bf 51}, 605 (1983).  \\
\end{document}